\documentclass[12pt]{article}
\usepackage{amsmath,url}
\usepackage{graphicx}
\usepackage{cite,hyperref}
\usepackage{subcaption}
\usepackage{multirow}
\providecommand{\href}[2]{#2}
 
\def\ltap{\raisebox{-.6ex}{\rlap{$\,\sim\,$}} \raisebox{.4ex}{$\,<\,$}}

\newcommand\as{\alpha_{\mathrm{S}}}

\def\to{\rightarrow}

\newcommand\Matrix{{\sc Matrix}}
\newcommand\Munich{{\sc Munich}}
\newcommand\OpenLoops{{\sc OpenLoops}}
\newcommand\Collier{{\sc Collier}}
\newcommand{\CutTools}{{\sc CutTools}}
\newcommand{\OneLOop}{{\sc OneLOop}}

\newcommand{\zz}{\ensuremath{ZZ}}
\newcommand{\z}{\ensuremath{Z}}

\newcommand{\abbrev}{}

\newcommand{\lo}{\text{\abbrev LO}}
\newcommand{\nlo}{\text{\abbrev NLO}}
\newcommand{\nnlo}{\text{\abbrev NNLO}}

\newcommand{\qcd}{{\abbrev QCD}}

\newcommand{\citere}[1]{Ref.~\cite{#1}}
\newcommand{\citeres}[1]{Refs.~\cite{#1}}
\newcommand{\reffi}[1]{Fig.~\ref{#1}}

\newcommand{\refta}[1]{Tab.~\ref{#1}}

\newcommand\Tstrut{\rule{0pt}{3.0ex}}         
\newcommand\Bstrut{\rule[-1.5ex]{0pt}{0pt}}   

\begin{document} 
\begin{titlepage}
\renewcommand{\thefootnote}{\fnsymbol{footnote}}
\begin{flushright}
ZU-TH 25/15\\
MITP/15-056
\end{flushright}
\vspace*{2cm}

\begin{center}
{\Large \bf \zz{} production at the LHC: fiducial cross sections\\[0.2cm] and distributions in \nnlo{} \qcd{}}
\end{center}

\par \vspace{2mm}
\begin{center}
{\bf Massimiliano Grazzini$^{(a)}$\footnote{On leave of absence from INFN, Sezione di Firenze, Sesto Fiorentino, Florence, Italy.}},
{\bf Stefan Kallweit$^{(b)}$} and {\bf Dirk Rathlev$^{(a)}$}
\vspace{5mm}

$^{(a)}$Physik-Institut, Universit\"at Z\"urich, CH-8057 Z\"urich, Switzerland 

$^{(b)}$PRISMA Cluster of Excellence, Institute of Physics,\\[0.1cm]
Johannes Gutenberg University, D-55099 Mainz, Germany

\vspace{5mm}

\end{center}

\par \vspace{2mm}
\begin{center} {\large \bf Abstract} \end{center}
\begin{quote}
\pretolerance 10000

We consider \qcd{} radiative corrections to the production of four charged leptons in 
the \zz{} signal region at the LHC.
We report on the complete calculation of the next-to-next-to-leading order (\nnlo{})
corrections to this process in \qcd{} perturbation theory.
Numerical results are presented for $\sqrt{s}=8$ TeV, using typical 
selection cuts applied by the ATLAS and CMS collaborations.
The \nnlo{} corrections increase the \nlo{} fiducial cross section by about $15\%$, and they have 
a relatively small impact on the shape
of the considered kinematical distributions. In the case of the $\Delta\Phi$ distribution of the two 
\z{} candidates, the \nnlo{} corrections improve the agreement of the theoretical prediction 
with the CMS data.

\end{quote}

\vspace*{\fill}
\begin{flushleft}
July 2015

\end{flushleft}
\end{titlepage}

\setcounter{footnote}{1}
\renewcommand{\thefootnote}{\fnsymbol{footnote}}

The production of \z{}-boson pairs at the Large Hadron Collider (LHC)
provides an important test of the electroweak (EW) sector of
the Standard Model (SM) at the TeV scale.
Small deviations in the observed rates or in the kinematical distributions could be a signal 
of new physics, possibly in the form of anomalous couplings.
At the same time, \zz{} production is an irreducible background
for Higgs boson production and new-physics searches.
Particularly important
are the off-shell effects below
the \zz{} threshold, relevant for the Higgs signal region in the four-lepton channel.
Various measurements of \zz{} hadroproduction have been carried out at the Tevatron and the LHC
(for some recent results see Refs.~\cite{CDF:2011ab,Abazov:2012cj,Aad:2012awa,Chatrchyan:2012sga,ATLAS-CONF-2013-020,Chatrchyan:2013oev,CMS:2014xja}).

From the theory side the first \nlo{} predictions for on-shell \zz{} production
were obtained long ago~\cite{Ohnemus:1990za,Mele:1990bq}.
The leptonic decays of the \z{} bosons
were included, initially neglecting spin correlations in the virtual 
contributions~\cite{Ohnemus:1994ff}.
The computation of the relevant one-loop helicity amplitudes~\cite{Dixon:1998py} enabled
the first complete \nlo{} calculations~\cite{Campbell:1999ah,Dixon:1999di}, including
spin correlations and off-shell effects.
The loop-induced gluon-fusion production channel, which formally contributes only at the
next-to-next-to-leading order (\nnlo{}),
was computed in \citeres{Glover:1988rg,Dicus:1987dj}. 
The corresponding leptonic decays were taken into account
in \citeres{Matsuura:1991pj,Zecher:1994kb,Binoth:2008pr}.
\nlo{} predictions for \zz{} production including the gluon-induced contribution,
the leptonic decays with spin correlations and off-shell effects were presented in \citere{Campbell:2011bn}.
The \nlo{} \qcd{} corrections to on-shell $ZZ+{\rm jet}$ production were discussed in 
\citere{Binoth:2009wk,Binoth:2010ra}, and the EW corrections to \zz{} production in 
\citere{Bierweiler:2013dja}.
A decisive step forward was carried out in \citere{Cascioli:2014yka}
where the inclusive \nnlo{} cross section for on-shell \zz{} production was presented.
This calculation was based on the evaluation of the two-loop amplitude for
on-shell \zz{} production. 
Later, the two-loop helicity amplitudes for all the vector-boson pair production processes 
were presented~\cite{Caola:2014iua,Gehrmann:2015ora}. This computation
paves the way to the consistent inclusion of the leptonic decays and off-shell effects in 
the \nnlo{} computation.

In this Letter we carry out this step
by considering \zz{} production at \nnlo{} including the leptonic decays of the vector bosons 
together with
spin correlations and off-shell effects. 
Contributions from $Z\gamma^*$ and $\gamma^*\gamma^*$ production as well as from
$pp\to Z/\gamma^*\to4\textrm{ leptons}$ topologies are also
consistently included with all interference terms.
Our calculation allows us to apply
arbitrary cuts on the final-state leptons and
the associated \qcd{} radiation.
Here we present selected numerical results
for $pp\to 4$ leptons at the LHC in \nnlo{} \qcd{},
using the typical cuts that are applied in the experimental \zz{} analyses.

Our calculation is performed with the numerical program
\Matrix\footnote{\Matrix{} is the abbreviation of 
``\Munich{} Automates qT subtraction and Resummation
to Integrate Cross Sections'', by M.~Grazzini, S.~Kallweit, D.~Rathlev, M.~Wiesemann. 
In preparation.}, which combines the $q_T$ subtraction~\cite{Catani:2007vq} and 
resummation~\cite{Bozzi:2005wk} formalisms with the 
\Munich{} Monte Carlo framework~\cite{Kallweit:Munich}.
\Munich{} provides a fully automated implementation of the Catani--Seymour dipole 
subtraction method~\cite{Catani:1996jh,Catani:1996vz},
an efficient phase-space integration,
as well as an interface to the one-loop generator \OpenLoops{}~\cite{Cascioli:2011va} 
to obtain all required (spin- and colour-correlated) 
tree-level and one-loop amplitudes.
For the numerically stable evaluation of tensor integrals, \OpenLoops{}
relies on the \Collier{} library~\cite{Denner:2014gla}, which is based on the 
Denner--Dittmaier reduction techniques~\cite{Denner:2002ii,Denner:2005nn} and the scalar 
integrals of~\cite{Denner:2010tr}.
To deal with problematic phase-space points, a rescue system is provided,
which employs the quadruple-precision implementation of the OPP method in 
\CutTools{}~\cite{Ossola:2007ax} and scalar integrals from 
\OneLOop{}~\cite{vanHameren:2010cp}.
Our implementation of $q_T$ subtraction and resummation for the production of colourless final states
is fully general, and it is based on the universality
of the hard-collinear coefficients~\cite{Catani:2013tia} appearing in transverse-momentum 
resummation.
These coefficients were explicitly computed for quark-initiated processes
in \citeres{Catani:2012qa,Gehrmann:2012ze,Gehrmann:2014yya}.
For the two-loop helicity amplitudes we use the results of \citere{Gehrmann:2015ora},
and of \citere{Matsuura:1988sm} for Drell--Yan like topologies.

A preliminary version of \Matrix{} has been employed in the \nnlo{} computations of
\citeres{Grazzini:2013bna,Cascioli:2014yka,Gehrmann:2014fva,Grazzini:2015nwa}, and in the 
resummed calculation of \citere{Grazzini:2015wpa}.

We consider $pp$ collisions with $\sqrt{s}=8$ TeV.
As for the EW couplings, we use the so-called $G_\mu$ scheme,
where the input parameters are $G_F$, $m_W$, $m_Z$.
More precisely, consistent with the \OpenLoops{} implementation, we use the complex $W$ 
and \z{} boson masses
to define the EW mixing angle as 
$\cos\theta_W^2=(m_W^2-i\Gamma_W\,m_W)/(m_Z^2-i\Gamma_Z\,m_Z)$.
In particular, we use the values
$G_F = 1.16639\times 10^{-5}$~GeV$^{-2}$, $m_W=80.399$ GeV, $\Gamma_W=2.1054$ GeV,
$m_Z = 91.1876$~GeV, $\Gamma_Z=2.4952$ GeV.
For the top quark we use $m_t=173.2$ GeV, $\Gamma_t=1.4426$ GeV, and for the Higgs 
boson $m_H=125$ GeV, $\Gamma_H=4.07$ MeV. Both the top quark and the Higgs boson only appear 
in diagrams with closed top-quark loops, thus entering the gluon-fusion channel and the real--virtual 
contribution.%
\footnote{
The Higgs boson contributes less than $1\%$ 
to the loop-induced $gg\to ZZ$ cross section, whereas its effect on the real--virtual contribution
is numerically negligible.}
We use the NNPDF3.0~\cite{Ball:2014uwa} sets of parton distributions with $\as(m_Z)=0.118$,
and the $\as$ running is evaluated at each corresponding order
(i.e., we use $(n+1)$-loop $\as$ at N$^n$LO, with $n=0,1,2$).
We consider $N_f=5$ massless quark flavours. The central
renormalization ($\mu_R$) and factorization ($\mu_F$) scales are set to
$\mu_R=\mu_F=m_Z$.

We first consider the ATLAS analysis of \citere{ATLAS-CONF-2013-020} in the three 
decay channels $e^+e^-e^+e^-$, $\mu^+\mu^-\mu^+\mu^-$, and $e^+e^-\mu^+\mu^-$.
The invariant masses of the two reconstructed lepton pairs
are required to fullfil the condition 66~GeV $\leq m_{ll}\leq$~116 GeV.
In the case of two lepton pairs with the same flavours there is a pairing ambiguity, 
which is resolved
by choosing the pairing that makes the sum of the absolute distances from $m_Z$ smaller.
The leptons are required to have $p_T\geq 7$~GeV and rapidity $|\eta|\leq 2.7$. For any 
lepton pair we require $\Delta R(l,l^\prime)>0.2$, independently of the flavours and charges
of $l$ and $l'$.

The corresponding cross sections are reported in \refta{tableatlas}, where the ATLAS 
results are also shown. The uncertainties on our theoretical predictions are obtained by 
varying the renormalization and factorization scales
in the range $0.5 m_Z<\mu_R,\mu_F<2m_Z$ with the constraint $0.5<\mu_F/\mu_R<2$.
Independently of the leptonic decay channels, 
the \nnlo{} corrections increase the \nlo{} result by about $15\%$, similarly to what was found for the 
inclusive cross section for on-shell \zz{} production~\cite{Cascioli:2014yka}.
This is as expected because the selection cuts are mild and do not significantly alter
the impact of radiative corrections.
The scale uncertainties are about $\pm 3\%$ at \nlo{} and remain of the same order at \nnlo{}.
As noted for the inclusive cross section~\cite{Cascioli:2014yka}, the \nlo{} scale uncertainty 
does not cover the \nnlo{} effect.
This is not surprizing since the loop-induced gluon-fusion contribution, which provides 
$\sim60\%$ of the $\mathcal{O}\left(\as^2\right)$ correction, opens up only at \nnlo{}.
The \nnlo{} corrections improve the agreement of the theoretical prediction with the data for 
the $e^+e^-\mu^+\mu^-$ channel, whereas they deteriorate the agreement in the case of the $4e$ 
and $4\mu$ channels. We note, however, that the predicted fiducial cross sections are 
still consistent with the ATLAS measurements at the $1\sigma$ level within the 
statistics-dominated uncertainties. 

\renewcommand{\baselinestretch}{1.5}
\begin{table}[ht]
\begin{center}
\begin{tabular}{|c| c| c| c| c|}
\hline
Channel & $\sigma_{\textrm{LO}}$ (fb) & $\sigma_{\textrm{NLO}}$ (fb) & $\sigma_{\textrm{NNLO}}$ (fb) & $\sigma_{\textrm{exp}}$ (fb) \\ [0.5ex]
\hline
\Tstrut
$e^+e^-e^+e^-$ & \multirow{2}{*}{$3.547(1)^{+2.9\%}_{-3.9\%}$} & \multirow{2}{*}{$5.047(1)^{+2.8\%}_{-2.3\%}$} & \multirow{2}{*}{$5.79(2)^{+3.4\%}_{-2.6\%}$} & $4.6^{+0.8}_{-0.7}{\rm (stat)}^{+0.4}_{-0.4}{\rm (syst.)}^{+0.1}_{-0.1}{\rm (lumi.)}$\Bstrut\\
\cline{1-1}
\cline{5-5}
\Tstrut
$\mu^+\mu^-\mu^+\mu^-$ &  &  &  & $5.0^{+0.6}_{-0.5}{\rm (stat)}^{+0.2}_{-0.2}{\rm (syst.)}^{+0.2}_{-0.2}{\rm (lumi.)}$\Bstrut\\
\hline
\Tstrut
 $e^+e^-\mu^+\mu^-$ & $6.950(1)^{+2.9\%}_{-3.9\%}$ & $9.864(2)^{+2.8\%}_{-2.3\%}$ & $11.31(2)^{+3.2\%}_{-2.5\%}$ & $11.1^{+1.0}_{-0.9}{\rm (stat)}^{+0.5}_{-0.5}{\rm (syst.)}^{+0.3}_{-0.3}{\rm (lumi.)}$\Bstrut\\
\hline
\end{tabular}
\end{center}
\renewcommand{\baselinestretch}{1.0}
\caption{\label{tableatlas} Fiducial cross sections and scale uncertainties for ATLAS cuts 
at \lo{}, \nlo{}, and \nnlo{} in the three considered leptonic decay channels. The ATLAS data are also shown.}
\end{table}

Secondly, we consider the CMS analysis of \citere{CMS:2014xja}. The fiducial region is defined 
as follows: all muons are required to fulfill $p_T^\mu>5$ GeV, $|\eta^\mu|<2.4$, while all 
electrons are required to fulfill $p_T^e>7$~GeV, $|\eta^e|<2.5$. In addition, the leading- 
and subleading-lepton transverse momenta must satisfy $p_T^{l,1}>20$ GeV and $p_T^{l,2}>10$ GeV, 
respectively. 
In the case of two lepton pairs with the same flavours, the pairing ambiguity is 
resolved by choosing the pair with the smallest distance from $m_Z$. 
This pair is called $Z_1$, the remaining pair is called $Z_2$. 
The invariant masses of the two reconstructed lepton pairs
are required to fulfill 60 GeV $\leq m_{ll}\leq$ 120 GeV. We note that in the case of identical 
flavours this definition of the fiducial region does not prevent the invariant masses of the
other two possible lepton pairs from becoming arbitrarily small,
giving rise to a collinear $\gamma^\ast\to l^-l^+$ singularity.
To avoid that, we follow CMS and
add an additional cut $m_{ll}>4$ GeV on all lepton pairs of the same flavours and opposite charges.\footnote{We thank Alexander Savin for providing us with this information.}
The corresponding fiducial cross sections and scale uncertainties, computed as above,
are reported in \refta{tablecms}.
Like for the ATLAS analysis, the \nnlo{} corrections increase the \nlo{} 
fiducial cross section
by about $15\%$. The scale uncertainties are similar to those reported in 
\refta{tableatlas}.

\renewcommand{\baselinestretch}{1.5}
\begin{table}[ht]
\begin{center}
\begin{tabular}{|c| c| c| c|}
\hline
Channel & $\sigma_{\textrm{LO}}$ (fb) & $\sigma_{\textrm{NLO}}$ (fb) & $\sigma_{\textrm{NNLO}}$ (fb) \\ [0.5ex]
\hline
\Tstrut
$e^+e^-e^+e^-$ & $3.149(1)^{+3.0\%}_{-4.0\%}$ & $4.493(1)^{+2.8\%}_{-2.3\%}$ & $5.16(1)^{+3.3\%}_{-2.6\%}$ \Bstrut\\
\hline
\Tstrut
$\mu^+\mu^-\mu^+\mu^-$ & $2.973(1)^{+3.1\%}_{-4.1\%}$ & $4.255(1)^{+2.8\%}_{-2.3\%}$ & $4.90(1)^{+3.4\%}_{-2.6\%}$\Bstrut\\
\hline
\Tstrut
 $e^+e^-\mu^+\mu^-$ & $6.179(1)^{+3.1\%}_{-4.0\%}$ & $8.822(1)^{+2.8\%}_{-2.3\%}$ & $10.15(2)^{+3.3\%}_{-2.6\%}$ \Bstrut\\
\hline
\end{tabular}
\end{center}
\renewcommand{\baselinestretch}{1.0}
\caption{\label{tablecms}
Fiducial cross sections and scale uncertainties for CMS cuts at \lo{}, \nlo{}, and \nnlo{} in the 
three considered leptonic decay channels.}
\end{table}

\begin{figure}[htpb]
        \centering
        \includegraphics[width=0.8\textwidth]{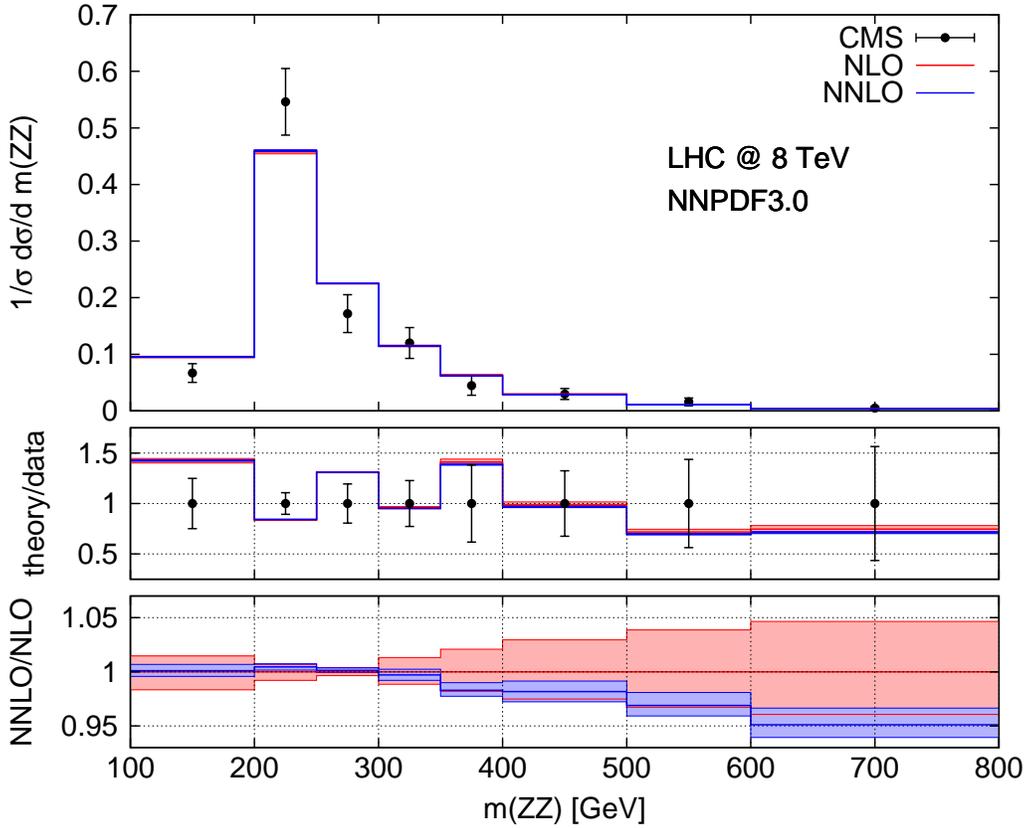}
    \caption{The four-lepton invariant-mass distribution
    at \nlo{} and \nnlo{} compared to the CMS data. In the lower panels the ratio of our theoretical results 
over the data, and the \nnlo{} result normalized to the central \nlo{} prediction are presented. The bands correspond to scale variations as described in the text.}
    \label{fig:1}
\end{figure}

CMS does not report the fiducial cross sections corresponding to the above cuts, but only 
normalized distributions, to which we compare our results.
We start with the invariant-mass distribution of the four leptons, which is depicted 
in \reffi{fig:1}. The lower panels show the theory/data comparison, and the \nnlo{} result normalized 
to the central \nlo{} prediction. We see that the \nnlo{} corrections have a limited impact in the 
comparison with the data, which still have large uncertainties.
The \nnlo{} effects on the normalized distribution are relatively small: they are completely negligible 
at low invariant masses, and they increase to $-5\%$ in the high mass region. 
This means that the \nnlo{} corrections make
the invariant mass distribution slightly softer. We have checked that this effect is due to 
the gluon-fusion contribution, whose relative effect decreases at high masses, due to the 
larger values of Bjorken $x$ that are probed.
The \nlo{} (\nnlo{}) scale uncertainties range from about $\pm 2\%$ ($\pm 1\%$) at low $m_{ZZ}$
to  $\pm 4\%$ ($\pm 2\%$) at high $m_{ZZ}$.

\begin{figure}[htpb]
    \begin{subfigure}[b]{0.49\textwidth}
        \centering
        \includegraphics[width=\textwidth]{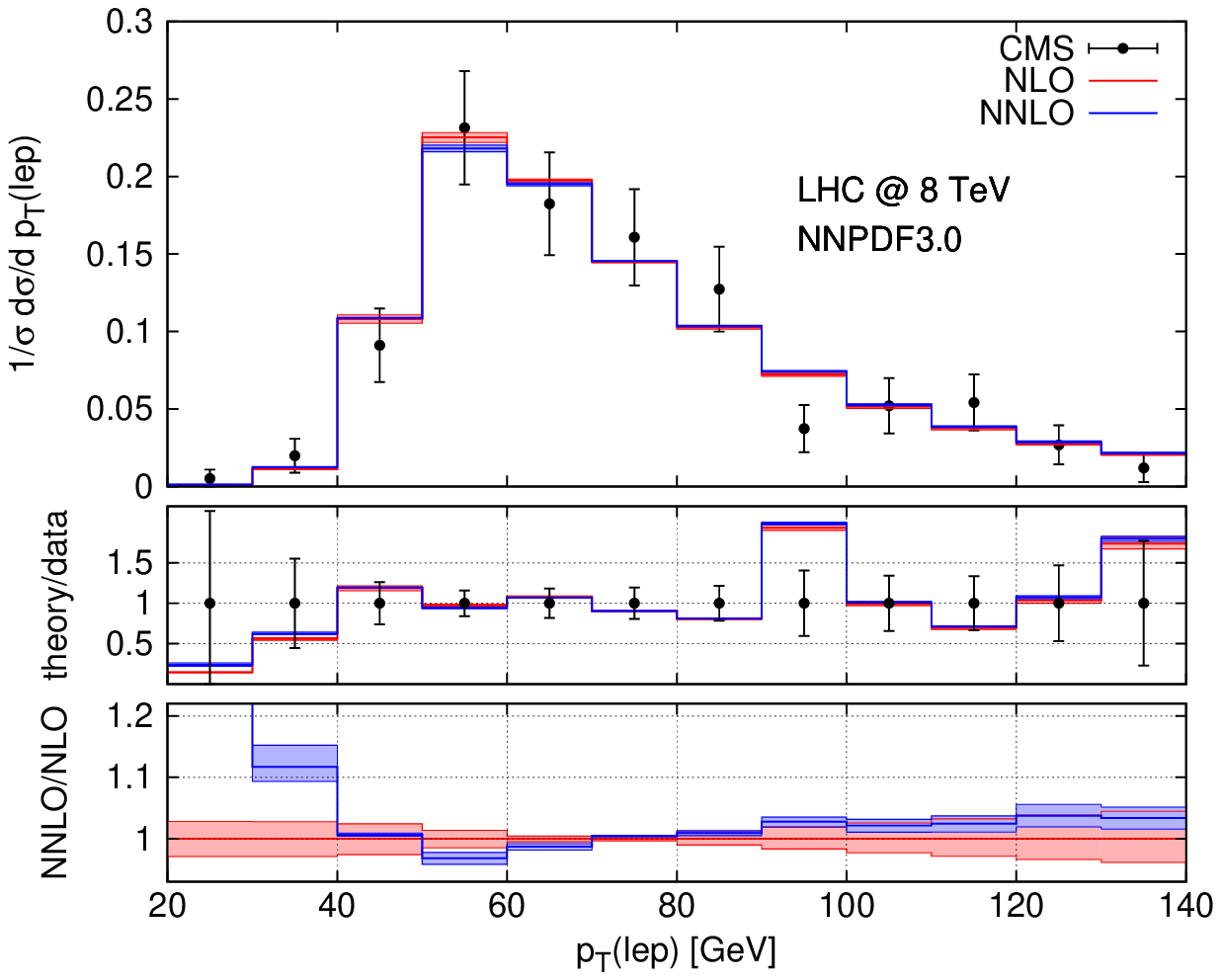}
    \end{subfigure}
    \begin{subfigure}[b]{0.49\textwidth}
        \centering
        \includegraphics[width=\textwidth]{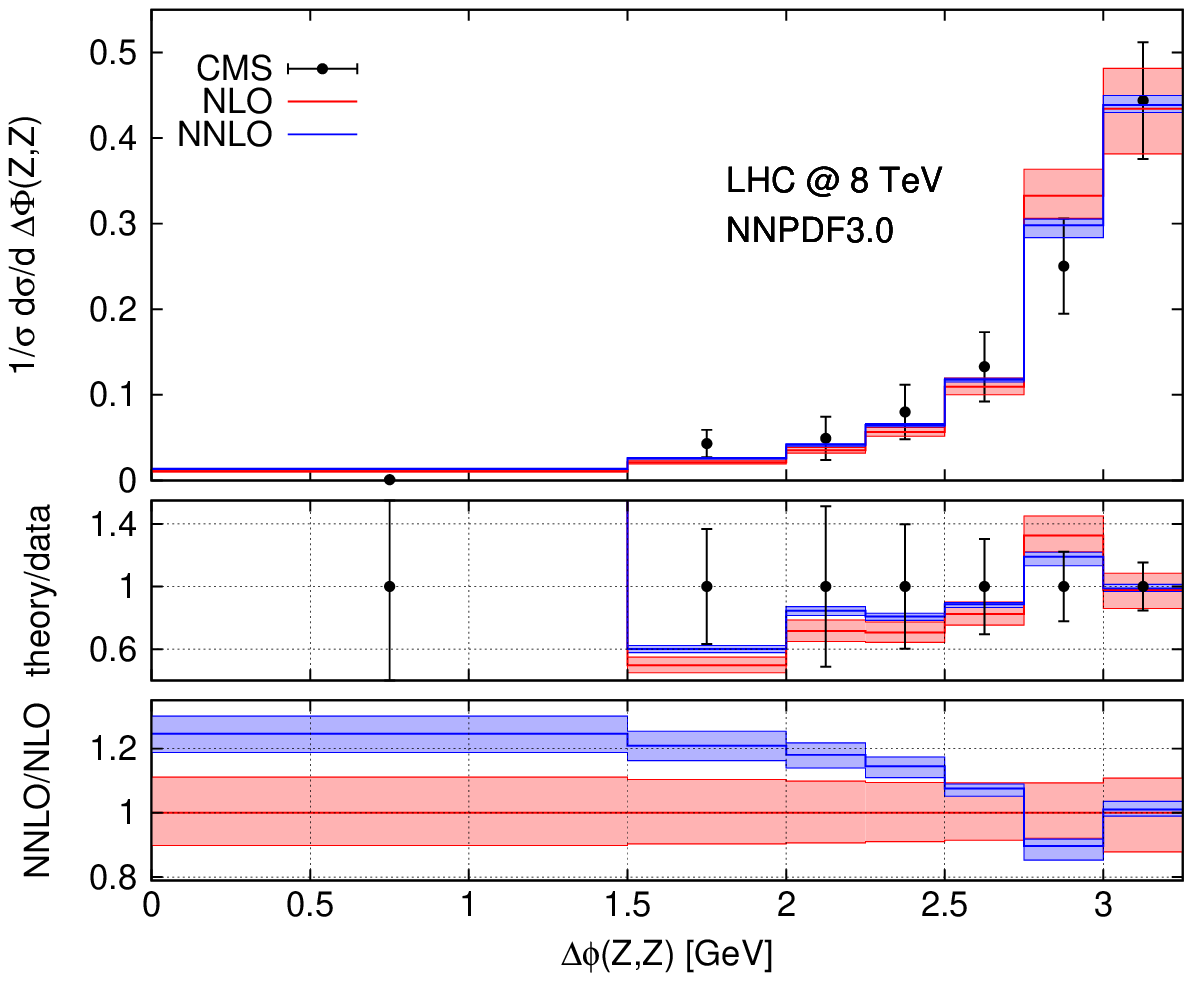}
    \end{subfigure}
    \caption{The leading-lepton $p_T$ (left) and the $\Delta\phi$ (right) distributions
    at \nlo{} and \nnlo{} compared to the CMS data.
In the lower panels the ratio of our theoretical results over the data, and the \nnlo{} result normalized to the central \nlo{} prediction are presented. The bands correspond to scale variations as described in the text.}
    \label{fig:2}
\end{figure}

In \reffi{fig:2} we show the analogous results for the leading-lepton $p_T$ distribution (left)
and the azimuthal separation ($\Delta\Phi$) of the two \z{} candidates (right). As in \reffi{fig:1}, 
we see that the \nnlo{} effects on the $p_T$ distribution
do not change the comparison with the data in a significant way.
The \nnlo{} corrections are also relatively small in most of the range of $p_T$ considered, 
except
for the low $p_T$ region, where they increase significantly. This effect is due to the 
gluon-fusion contribution, whose relative impact increases as $p_T$ decreases.
The situation is different for the $\Delta\Phi$ distribution. Here the \nnlo{} corrections 
improve the agreement with the data,
except for the first bin, where the CMS measurement is an order of magnitude below the 
theoretical \nnlo{} prediction. The larger impact of \nnlo{} corrections in the $\Delta\Phi$ 
distribution can
be understood easily by the observation that at \lo{} the reconstructed \z{} bosons are always
back-to-back, i.e., $\Delta\Phi(Z_1,Z_2)=\pi$. As a consequence, the \nnlo{} calculation is 
effectively \nlo{} in the region $0\leq\Delta\Phi<\pi$. The \nnlo{} corrections amount to 
about $+25\%$ when $\Delta\Phi\ltap 1.5$, and decrease as $\Delta\Phi$ increases. 
We note that this effect is entirely due to the \nnlo{} corrections to the $q{\bar q}$ channel 
addressed in this paper, since the loop-induced gluon-fusion contribution, which also enters at \nnlo{}, 
affects the $\Delta\Phi$ distribution only at $\Delta\Phi=\pi$.
The \nlo{} scale uncertainties are about $\pm 11\%$, while at \nnlo{} the uncertainties are 
about $\pm 5\%$ at low $\Delta\Phi$, and decrease to about $\pm 2\%$ at high $\Delta\Phi$.

We have presented the first complete NNLO QCD calculation for the production of four charged leptons in
the \zz{} signal region at the LHC. We have studied the impact of \nnlo{} corrections on the fiducial cross 
sections and distributions measured by ATLAS and CMS at the LHC. 
As for the fiducial cross 
sections, we found about $+15\%$ \nnlo{} corrections w.r.t.\ the \nlo{} prediction, consistent with what was found 
for the inclusive cross section for on-shell \zz{} production~\cite{Cascioli:2014yka}. 
The impact on the normalized 
distributions we considered is small compared to the experimental uncertainties, but 
leads to an improved agreement with the data in the case
of the $\Delta\Phi$ distribution of the two \z{} candidates. Our calculation was performed
with the numerical program \Matrix, which is able to carry out fully exclusive \nnlo{} computations for a 
wide class of processes at hadron colliders. We look forward to further applications of our 
framework
to other important LHC processes.

\noindent {\bf Acknowledgements.}
This research was supported in part by the Swiss National Science Foundation (SNF) under 
contracts CRSII2-141847, 200021-156585, and by 
the Research Executive Agency (REA) of the European Union under the Grant Agreement 
number PITN--GA--2012--316704 ({\it HiggsTools}).


\begin{thebibliography}{99}




\bibitem{CDF:2011ab}
  T.~Aaltonen {\it et al.}  [CDF Collaboration],
  Phys.\ Rev.\ Lett.\  {\bf 108} (2012) 101801
  [arXiv:1112.2978 [hep-ex]].


\bibitem{Abazov:2012cj}
  V.~M.~Abazov {\it et al.}  [D0 Collaboration],
  Phys.\ Rev.\ D {\bf 85} (2012) 112005
  [arXiv:1201.5652 [hep-ex]].


\bibitem{Aad:2012awa}
  G.~Aad {\it et al.}  [ATLAS Collaboration],
  JHEP {\bf 1303} (2013) 128
  [arXiv:1211.6096 [hep-ex]].




\bibitem{Chatrchyan:2012sga}
  S.~Chatrchyan {\it et al.}  [CMS Collaboration],
  JHEP {\bf 1301} (2013) 063
  [arXiv:1211.4890 [hep-ex]].

\bibitem{ATLAS-CONF-2013-020}
ATLAS Collaboration, ATLAS-CONF-2013-020.


\bibitem{Chatrchyan:2013oev}
  S.~Chatrchyan {\it et al.}  [CMS Collaboration],
  Phys.\ Lett.\ B {\bf 721} (2013) 190
  [arXiv:1301.4698 [hep-ex]].

\bibitem{CMS:2014xja}
  V.~Khachatryan {\it et al.}  [CMS Collaboration],
  Phys.\ Lett.\ B {\bf 740} (2015) 250
  [arXiv:1406.0113 [hep-ex]].


\bibitem{Ohnemus:1990za}
  J.~Ohnemus and J.~F.~Owens,
  Phys.\ Rev.\ D {\bf 43} (1991) 3626.

\bibitem{Mele:1990bq}
  B.~Mele, P.~Nason and G.~Ridolfi,
  Nucl.\ Phys.\ B {\bf 357} (1991) 409.


\bibitem{Ohnemus:1994ff}
  J.~Ohnemus,
  Phys.\ Rev.\ D {\bf 50} (1994) 1931
  [hep-ph/9403331].


\bibitem{Dixon:1998py}
  L.~J.~Dixon, Z.~Kunszt and A.~Signer,
  Nucl.\ Phys.\ B {\bf 531} (1998) 3
  [hep-ph/9803250].

\bibitem{Campbell:1999ah}
  J.~M.~Campbell and R.~K.~Ellis,
  Phys.\ Rev.\ D {\bf 60} (1999) 113006
  [hep-ph/9905386].


\bibitem{Dixon:1999di}
  L.~J.~Dixon, Z.~Kunszt and A.~Signer,
  Phys.\ Rev.\ D {\bf 60} (1999) 114037
  [hep-ph/9907305].


\bibitem{Glover:1988rg}
  E.~W.~N.~Glover and J.~J.~van der Bij,
  Nucl.\ Phys.\ B {\bf 321} (1989) 561.


\bibitem{Dicus:1987dj}
  D.~A.~Dicus, C.~Kao and W.~W.~Repko,
  Phys.\ Rev.\ D {\bf 36} (1987) 1570.


\bibitem{Matsuura:1991pj}
  T.~Matsuura and J.~J.~van der Bij,
  Z.\ Phys.\ C {\bf 51} (1991) 259.

\bibitem{Zecher:1994kb}
  C.~Zecher, T.~Matsuura and J.~J.~van der Bij,
  Z.\ Phys.\ C {\bf 64} (1994) 219
  [hep-ph/9404295].

\bibitem{Binoth:2008pr}
  T.~Binoth, N.~Kauer and P.~Mertsch,
  arXiv:0807.0024 [hep-ph].


\bibitem{Campbell:2011bn}
  J.~M.~Campbell, R.~K.~Ellis and C.~Williams,
  JHEP {\bf 1107} (2011) 018
  [arXiv:1105.0020 [hep-ph]].


\bibitem{Binoth:2009wk}
  T.~Binoth, T.~Gleisberg, S.~Karg, N.~Kauer and G.~Sanguinetti,
  Phys.\ Lett.\ B {\bf 683} (2010) 154
  [arXiv:0911.3181 [hep-ph]].

\bibitem{Binoth:2010ra}
  J.~R.~Andersen {\it et al.}  [SM and NLO Multileg Working Group Collaboration],
  arXiv:1003.1241 [hep-ph].


\bibitem{Bierweiler:2013dja}
  A.~Bierweiler, T.~Kasprzik and J.~H.~K\"uhn,
  JHEP {\bf 1312} (2013) 071
  [arXiv:1305.5402 [hep-ph]].


\bibitem{Cascioli:2014yka}
  F.~Cascioli, T.~Gehrmann, M.~Grazzini, S.~Kallweit, P.~Maierh\"ofer, A.~von Manteuffel, S.~Pozzorini, D.~Rathlev, L.~Tancredi and E.~Weihs,
  Phys.\ Lett.\ B {\bf 735} (2014) 311
  [arXiv:1405.2219 [hep-ph]].



\bibitem{Caola:2014iua}
  F.~Caola, J.~M.~Henn, K.~Melnikov, A.~V.~Smirnov and V.~A.~Smirnov,
  JHEP {\bf 1411} (2014) 041
  [arXiv:1408.6409 [hep-ph]].

\bibitem{Gehrmann:2015ora}
  T.~Gehrmann, A.~von Manteuffel and L.~Tancredi,
  arXiv:1503.04812 [hep-ph].


\bibitem{Catani:2007vq}
  S.~Catani and M.~Grazzini,
  Phys.\ Rev.\ Lett.\  {\bf 98} (2007) 222002
[hep-ph/0703012].



\bibitem{Bozzi:2005wk}
  G.~Bozzi, S.~Catani, D.~de Florian and M.~Grazzini,
  Nucl.\ Phys.\ B {\bf 737} (2006) 73
  [hep-ph/0508068].


\bibitem{Kallweit:Munich}
S.~Kallweit,
\Munich{} is the abbreviation of ``MUlti-chaNnel Integrator at Swiss~(CH) precision''---an automated parton level NLO 
generator. 
In preparation.


\bibitem{Catani:1996jh}
  S.~Catani and M.~H.~Seymour,
  Phys.\ Lett.\ B {\bf 378} (1996) 287
  [hep-ph/9602277].

\bibitem{Catani:1996vz}
  S.~Catani and M.~H.~Seymour,
  Nucl.\ Phys.\ B {\bf 485} (1997) 291
   [Erratum-ibid.\ B {\bf 510} (1998) 503]
[hep-ph/9605323].




\bibitem{Cascioli:2011va}
  F.~Cascioli, P.~Maierh\"ofer and S.~Pozzorini,
  Phys.\ Rev.\ Lett.\  {\bf 108} (2012) 111601
  [arXiv:1111.5206 [hep-ph]].

\bibitem{Denner:2014gla}
  A.~Denner, S.~Dittmaier and L.~Hofer,
  PoS LL {\bf 2014} (2014) 071
  [arXiv:1407.0087 [hep-ph]].



\bibitem{Denner:2002ii}
  A.~Denner and S.~Dittmaier,
  Nucl.\ Phys.\ B {\bf 658} (2003) 175
  [hep-ph/0212259].


\bibitem{Denner:2005nn}
  A.~Denner and S.~Dittmaier,
  Nucl.\ Phys.\ B {\bf 734} (2006) 62
  [hep-ph/0509141].

\bibitem{Denner:2010tr}
  A.~Denner and S.~Dittmaier,
  Nucl.\ Phys.\ B {\bf 844} (2011) 199
  [arXiv:1005.2076 [hep-ph]].

\bibitem{Ossola:2007ax}
  G.~Ossola, C.~G.~Papadopoulos and R.~Pittau,
  JHEP {\bf 0803} (2008) 042
  [arXiv:0711.3596 [hep-ph]].

\bibitem{vanHameren:2010cp}
  A.~van Hameren,
  Comput.\ Phys.\ Commun.\  {\bf 182} (2011) 2427
  [arXiv:1007.4716 [hep-ph]].

\bibitem{Catani:2013tia}
  S.~Catani, L.~Cieri, D.~de Florian, G.~Ferrera and M.~Grazzini,
  Nucl.\ Phys.\ B {\bf 881} (2014) 414
  [arXiv:1311.1654 [hep-ph]].


\bibitem{Catani:2012qa}
  S.~Catani, L.~Cieri, D.~de Florian, G.~Ferrera and M.~Grazzini,
  Eur.\ Phys.\ J.\ C {\bf 72} (2012) 2195
[arXiv:1209.0158 [hep-ph]].



\bibitem{Gehrmann:2012ze}
  T.~Gehrmann, T.~L\"ubbert and L.~L.~Yang,
  Phys.\ Rev.\ Lett.\  {\bf 109} (2012) 242003
[arXiv:1209.0682 [hep-ph]],

\bibitem{Gehrmann:2014yya}
  T.~Gehrmann, T.~L\"ubbert and L.~L.~Yang,
  JHEP {\bf 1406} (2014) 155
  [arXiv:1403.6451 [hep-ph]].

\bibitem{Matsuura:1988sm}
  T.~Matsuura, S.~C.~van der Marck and W.~L.~van Neerven,
  Nucl.\ Phys.\ B {\bf 319} (1989) 570.



\bibitem{Grazzini:2013bna}
  M.~Grazzini, S.~Kallweit, D.~Rathlev and A.~Torre,
  Phys.\ Lett.\ B {\bf 731} (2014) 204
  [arXiv:1309.7000 [hep-ph]].


\bibitem{Gehrmann:2014fva}
  T.~Gehrmann, M.~Grazzini, S.~Kallweit, P.~Maierh\"ofer, A.~von Manteuffel, S.~Pozzorini, D.~Rathlev and L.~Tancredi,
  Phys.\ Rev.\ Lett.\  {\bf 113} (2014) 21,  212001
  [arXiv:1408.5243 [hep-ph]].

\bibitem{Grazzini:2015nwa}
  M.~Grazzini, S.~Kallweit and D.~Rathlev,
  arXiv:1504.01330 [hep-ph].

\bibitem{Grazzini:2015wpa}
  M.~Grazzini, S.~Kallweit, D.~Rathlev and M.~Wiesemann,
  arXiv:1507.02565 [hep-ph].


\bibitem{Ball:2014uwa}
  R.~D.~Ball {\it et al.}  [NNPDF Collaboration],
  JHEP {\bf 1504} (2015) 040
  [arXiv:1410.8849 [hep-ph]].



\end{thebibliography}
\end{document}